\documentclass[twocolumn,prl,superscriptaddress]{revtex4}
\usepackage{graphics}
\usepackage{graphicx}
\usepackage{amsfonts}
\usepackage{textcomp}
\usepackage{amssymb}
\usepackage{mathrsfs}
\usepackage{amsmath}
\usepackage{color}
\usepackage{float}

\begin{document}

\title{Multihyperuniform Long-Range Order in Medium-Entropy Alloys}

\author{Duyu Chen}
\email[correspondence sent to: ]{duyu@alumni.princeton.edu}
\affiliation{Materials Research Laboratory, University of California, Santa Barbara, California 93106, United States}
\author{Xinyu Jiang}
\affiliation{Mechanical and Aerospace Engineering, Arizona State
University, Tempe, AZ 85287} 
\author{Duo Wang}
\affiliation{Materials Science and Engineering, Arizona State
University, Tempe, AZ 85287} 
\author{Houlong Zhuang}
\email[correspondence sent to: ]{hzhuang7@asu.edu}
\affiliation{Mechanical and Aerospace Engineering, Arizona State
University, Tempe, AZ 85287} 
\author{Yang Jiao}
\email[correspondence sent to: ]{yang.jiao.2@asu.edu}
\affiliation{Materials Science and Engineering, Arizona State
University, Tempe, AZ 85287} \affiliation{Department of Physics,
Arizona State University, Tempe, AZ 85287}


\begin{abstract}
We provide strong numerical evidence for a hidden multihyperuniform long-range order (MHLRO) in SiGeSn medium-entropy alloys (MEAs), in which the normalized infinite-wavelength composition fluctuations for all three atomic species are completely suppressed as in a perfect crystalline state. We show this MHLRO naturally leads to the emergence of short-range order (SRO) recently discovered in MEAs, which results in stable lower-energy states compared to alloy models with random or special quasi-random structures (SQSs) possessing no atomic SROs. The MHLRO MEAs approximately realize the Vegard’s law, which offers a rule-of-mixture type predictions of the lattice constants and electronic band gap, and thus can be considered as an ideal mixing state. The MHLRO also directly gives rise to enhanced electronic band gaps and superior thermal transport properties at low temperatures compared to random structures and SQSs, which open up  novel potential applications in optoelectronics and thermoelectrics. Our analysis of the SiGeSn system leads to the formulation of general organizing principles applicable in other medium- and high-entropy alloys (HEAs), and a highly efficient computational model for rendering realistic large-scale configurations of MEAs and HEAs.





\end{abstract}
\maketitle




Medium-entropy alloys (MEAs) are those composed of mixtures of equal or relatively large proportions of three elements, and alloys containing five or more elements are usually referred to as high-entropy alloys (HEAs) \cite{ZHANG20141}. These MEAs and HEAs are distinctly different from traditional metallic alloys which contain one or two major components with smaller amounts of other elements. Studies have shown that MEAs and HEAs can possess superior mechanical performance and resistance to corrosion compared to traditional alloys \cite{MIRACLE2017448}, and can have unusual thermal and electronic transport properties \cite{mu_pei_liu_stocks_2018}, opening up their new device applications. 


Atomic structures of medium/high entropy alloys largely remain a mystery, due to challenges in direct imaging of bulk structures in three dimensions (3D). The preponderance of previous studies typically assumed MEAs and HEAs possess a random or special quasi-random structure (SQS) \cite{PhysRevLett.65.353} consisting of atoms of different types distributed on the sites of a crystalline lattice. However, a number of recent studies, in particular those on the chromium-cobalt-nickel (CrCoNi) MEAs, have discovered significant short-range order, often manifested as suppressed clustering of atoms of the same type, which are shown to lead to energetically favorable states of the alloys \cite{ding2018tunable,zhang2020short, walsh2021magnetically}. It is interesting to note that a natural consequence of the existence of such short-range order is the suppression of composition (or density) fluctuations. Since the atoms prefer certain neighbor arrangements, it can be expected that as an observation window moves randomly within the disordered system, the number of atoms of a particular element that fall within the window would not fluctuate significantly, especially for large window sizes.







A system in which the infinite-wavelength normalized density fluctuations are completely suppressed is called {\it hyperuniform} \cite{To03}. Among the hyperuniform systems, the {\it disordered hyperuniform} (DHU) varieties are recently discovered exotic states of matter \cite{To03, To18a, Do05, Za11a, Ji11, Ji14} that lie between a perfect crystal and liquid. These systems are similar to liquids or glasses in that they are statistically isotropic and possess no Bragg peaks, and yet they completely suppress normalized large-scale density fluctuations like crystals and in this sense possess a \textit{hidden} long-range order \cite{To03, Za09, To18a}. Hyperuniformity is manifested as vanishing static structure factor $S(k)$ in the zero-wavenumber limit, i.e., $\lim_{k\rightarrow 0}S(k) = 0$ (where $k$ is the wavenumber), or equivalently vanishing normalized number variance infinite-wavelength limit, i.e., $\lim_{R\rightarrow\infty} \sigma^2_N(R)/N(R) = 0$, where $\sigma_N^2(R)$ and $N(R)$ are respectively the local number variance and average atom number in a spherical observation window of radius $R$ \cite{To03, To18a}. DHU states have been discovered in a variety of equilibrium and non-equilibrium physical and
biological systems, and appear to endow such systems with desirable physical properties that cannot be achieved in either the totally random or perfectly crystalline states \cite{Za11a, Ji11, Ch14, Za11b, To15, Do05, Ba09, Ku11, Hu122, Dr15, He15, Ja15, We15, Fl09, Ma13, Ji14, Ma15, He13, Kl19, Le19a, Le19b, Ch18a, Ch18b, Ru19, Hu21, To21, jiao2021hyperuniformity}.


We are thus motivated to ask the following fundamental question: Do MEAs (or HEAs) with significant short-range order also possess the aforementioned hidden hyperuniform long-range order? In this letter, using the ternary silicon-germanium-tin (SiGeSn) MEA as an example, we provide strong numerical evidence for a hidden {\it multihyperuniform} long-range order (MHLRO) in MEAs, in which the normalized infinite-wavelength composition fluctuations for {\it all} atomic species are completely suppressed as in a perfect crystalline state. This is achieved by numerically rendering multihyperuniform realizations of the alloy, which naturally leads to the emergence of short-range order (SRO) quantitatively consistent with recently reported observations and results in stable lower-energy states compared to alloys possessing random structure and SQS without any atomic SRO \cite{ding2018tunable,zhang2020short, walsh2021magnetically}. We note that multihyperuniform patterns have been discovered previously in other physical and biological systems, e.g., chicken eyes \cite{Ji14}.
The MHLRO MEAs approximately realize Vegard’s law \cite{vegard1921konstitution}, which assumes an ideal mixture, and predicts that for a multi-component solid solution where the individual component in its pure form shares the same crystal structure, the lattice parameter and electronic band gap of the mixture should be the weighted averages of the constituents' lattice parameters and band gaps. In particular, the MHLRO state minimizes the lattice distortion compared to the prediction of Vegard's law, and thus can be considered as an ideal mixing state. We also demonstrate that the MHLRO directly gives rise to enhanced electronic band gaps and superior thermal transport properties at low temperatures compared to the random/SQS alloy models, which open up many novel potential applications in optoelectronics and thermoelectrics. Based on our numerical analysis of SiGeSn system, we formulate general organizing principles predicting similar MHLRO in other MEAs and HEAs. 

{\bf Realizations and structural characteristics of multihyperuniform MEAs.} We devise a highly efficient general Fourier-space construction technique to generate large samples of disordered multihyperuniform MEAs (or HEAs) as well as those with SQSs. Specifically, we define a fictitious ``energy'' $E$ of the system as the squared differences between the target and constructed structure factors associated with different elements, i.e.,
\begin{equation}
\label{eq_1} E = \sum_{i}\sum_{k} [S_i(k)-S_{i,0}(k)]^2,
\end{equation}
where $S_{i,0}(k)$ and $S_i(k)$ are the angular-averaged structure factor associated with type-$i$ atoms in the target and constructed structures. We employ the simulated annealing procedure consisting of sequential Monte Carlo (MC) moves that involve swapping particles of different types to obtain low-energy states associated with this ``energy'' $E$, and drive the system under construction towards one with targeted structure factors (see Supplemental Materials for detail). 

To obtain multihyperuniform MEAs or HEAs, we impose the strong condition of stealthy hyperuniformity and require that the structure factor associated with each atomic species is zero for a range of small $k$, i.e., 
\begin{equation}
\label{eq_2} S_{i,0}(k)=0 ~~~~ \textnormal{for} ~~ k < K_{i,0},
\end{equation}
where $K_{i,0}$ is the range of exclusion region in the Fourier space for the target structure factor associated with element $i$. To obtain SQSs for MEAs or HEAs that faithfully retain the salient ensemble statistics of random mixtures, when considering type-$i$ atoms, we treat all the other types of atoms as ``voids'', and utilize a previously known formula for structure factor of lattice structures with voids to obtain the target structure factor associated with element $i$:
\begin{equation}
\label{eq_3} S_{i,0}(k)=p_i+(1-p_i)S_l(k)=1-c_i+c_iS_l(k), 
\end{equation}
where the ``void'' concentration $p_i=1-c_i$, $c_i$ is the atomic fraction of element $i$, and $S_l(k)$ is the structure factor associated with the underlying lattice structure that the MEA or HEA is based on.

In this letter, as an example we consider equal-concentration ternary SiGeSn alloys that were known to possess a diamond lattice structure, i.e., $c_\mathrm{Si}=c_\mathrm{Ge}=c_\mathrm{Sn}=1/3$. The SROs of such alloys have not been systematically explored before, and for simplicity we assume that $S_i(k)$ are the same for all three atom types, and set $K_{i,0}=K_0$ for all $i$ when generating multihyperuniform SiSeSn. This general setting also makes our structural model immediately applicable to other MEAs. To suppress the generation of periodic structures and at the same time ensure multihyperuniformity, we further pick $K_p < K_0 < K_1$, where $K_p$ is the wavenumber associated with the smallest possible periodic unit compatible with the prescribed atomic fractions, and $K_1$ is the location of the first Bragg peak associated with the underlying lattice (see Supplemental Materials for detailed explanation). We provide visualizations of representative SiGeSn structures with MHLRO in Fig. 1, and structure factors associated with different elements for $K_0a=3.6$ in Fig. 2, where $a$ is the lattice constant of a diamond cubic unit. As expected, the structure factor $S_{i}(k)$ decreases as $k$ decreases at small $k$ and essentially approaches zero as $k$ goes to zero, indicating hyperuniformity of the atomic distribution for all element types. However, the stealthiness is not realized, and $S_i(k)$ possesses small, but finite values at small $k<K_0$ in the exclusion region. The lack of stealthiness is likely due to the frustration caused by the presence of three element types as they simultaneously try to form a hyperuniform pattern \cite{Ji14}.

\begin{figure}[ht!]
\begin{center}
$\begin{array}{c}\\
\includegraphics[width=0.5\textwidth]{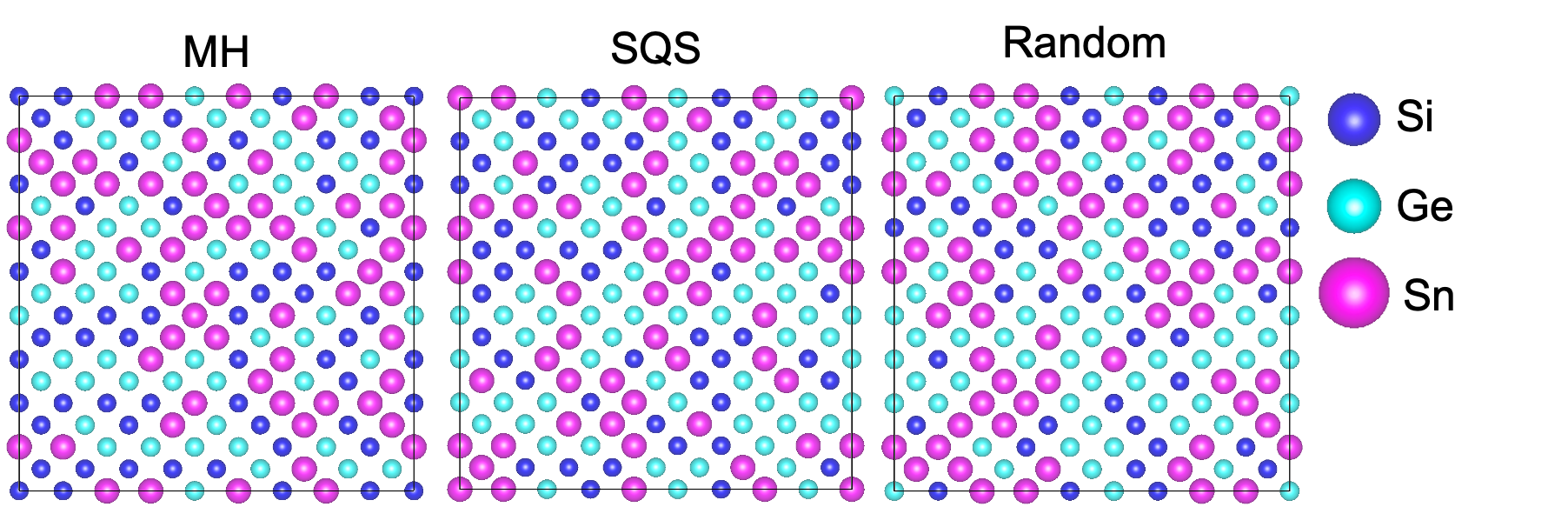} 
\end{array}$
\end{center}
\caption{(Color online) Two-dimensional slices along the (001) plane of representative multihyperuniform (with an exclusion region of $K_0a=3.6$ for targeted structure factor, where $a$ is the lattice constant of a diamond cubic unit), random, and special quasirandom structures of SiGeSn alloys on a 9~$\times$~9~$\times$~9 diamond lattice.} \label{fig_1}
\end{figure}

\begin{figure*}[t]
\begin{center}
\includegraphics[width=0.9\textwidth]{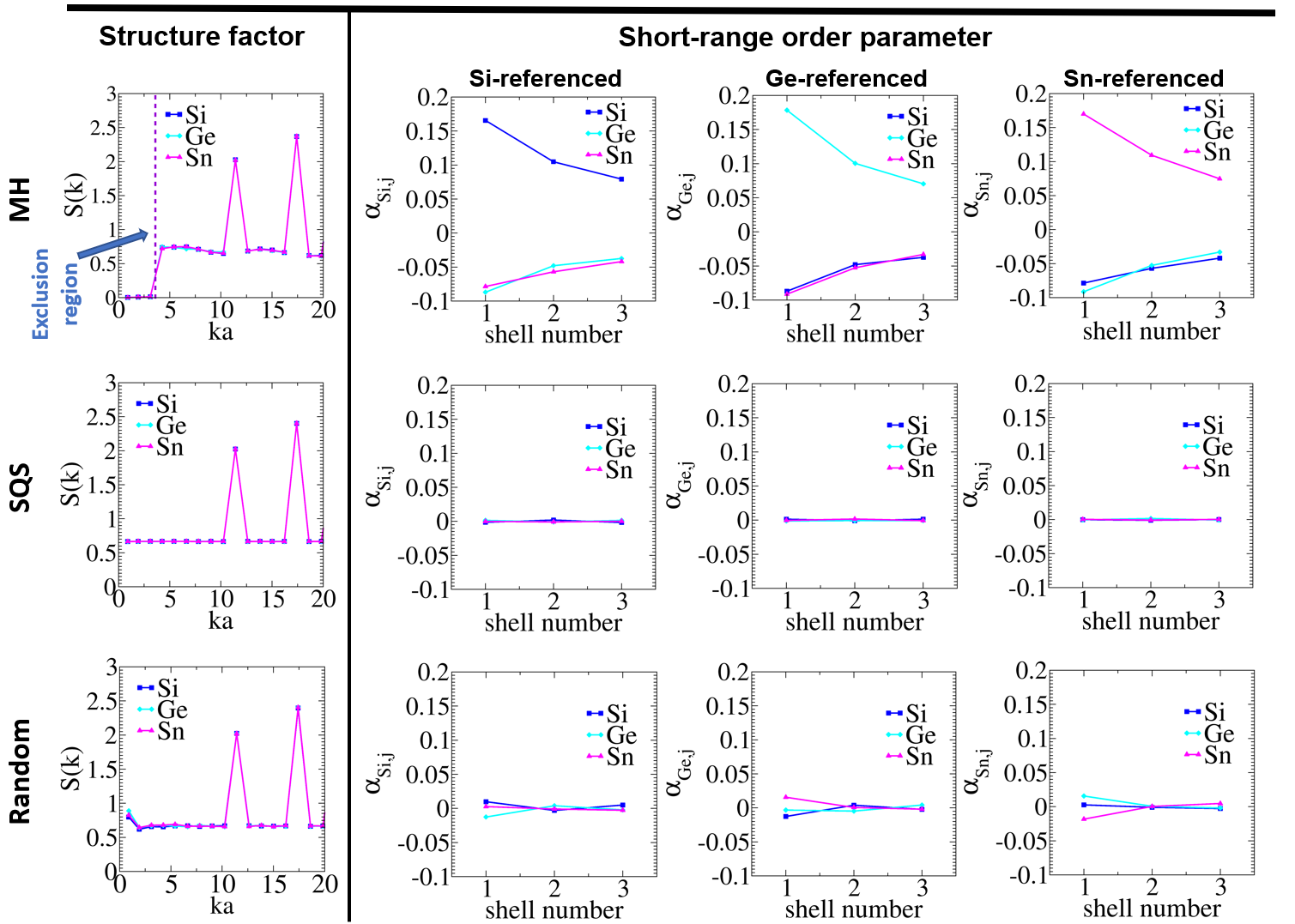} 
\end{center}
\caption{(Color online) Structure factors (left column) and Warren-Cowley short-range order parameters $\alpha_{ij}^{\nu}$ (right column) of multihyperuniform (with an exclusion region of $K_0a=3.6$ for targeted structure factor), random, and special quasirandom structures of SiGeSn alloys. The statistics are averaged over 5 configurations with $N=5832$ particles.} \label{fig_2}
\end{figure*}

As a comparison, we also generate corresponding random mixtures and SQSs for such alloys, as shown in Fig. 1. As expected, the structure factors $S_i(k)$ of random mixtures and SQSs are almost indistinguishable, and are essentially flat and proportional to $1-c_i$ away from the Bragg peaks, although the random mixtures possess larger statistical fluctuations, as shown in left column of Fig. 2. To quantify the SRO of multihyperuniform, random, and SQS alloys, we compute the Warren-Cowley SRO parameter $\alpha_{ij}^{\nu}$ \cite{Co50} defined as 
\begin{equation}
\label{eq_4} \alpha_{ij}^{\nu} = 1 - \frac{p_{ij}^{\nu}}{c_j},
\end{equation}
where $p_{ij}^{\nu}$ is the probability of finding atomic species $j$ around an atom of type $i$ in the $\nu$-th neighboring shell, and the results are shown in right column of Fig. 2. Clearly, the multihyperuniform structures possess substantial SRO, and atoms of the same type are unfavored in the first three shells around any atom in the system, while random and SQS structures possess essentially no SRO, which are expected. The natural emergence of SRO in MEAs with hidden MHLRO suggest that these two types of order may be coupled, and could be due to the fact that in order to achieve suppression of normalized composition fluctuations on large length scales for each species under lattice constraints, the system needs to suppress local clustering of the same type of atoms. This rationale is not unique to the SiGeSn alloy, and given the ubiquitous nature of SRO discovered previously in MEAs and HEAs, strongly indicates the possibility of hidden MHLRO in other MEAs or HEAs.

{\bf MHLRO leads to lower-energy states and smaller lattice distortions in MEAs.} Conventional Si-Ge-Sn alloys were believed to possess a single solid solution phase with amorphous atomic structures, consistent with the absence of any binary or ternary intermetallic compounds in the Si-Ge-Sn phase diagrams \cite{olesinski1984si,olesinski1984ge1,olesinski1984ge2}. However, it remains a long-standing question whether any sort of atomic ordering exists in Si-Ge-Sn alloys \cite{PhysRevB.95.161402}. For example, short-ranger order was recently predicted in Ge-Sn alloys \cite{cao2020short}, while previous Monte Carlo simulations that only explore the vicinity of random mixtures/SQSs in the configuration space fail to discover low-energy states with short-range order \cite{doi:10.1063/1.5135324}. Here we address this enduring issue by comparing the energies of our DHU structures with those of random and SQS systems. In particular, we perform molecular dynamics simulations with the Stillinger-Weber potential parameterized for Si-Ge-Sn alloys \cite{PhysRevB.31.5262,lee2012force,lee2017molecular}, which has been used to study thermal transport of these alloys \cite{WANG2020443}, and compute the energy difference using the DHU system's energy as the reference. We perform calculations for different supercell sizes ($N=5832$ and $N= 46656$ atoms), and benchmark the energy differences with the density functional theory (see the Supplemental Material for calculation parameters) results of systems with $N=216$ atoms. Figure 3(a) shows that although the magnitudes of the energy differences are both method and size-dependent, all the results consistently show that (i) the energies of the SQS and random systems are similar; more importantly, (ii) the multihyperuniform system is the most energetically stable, which strongly suggests the existence of long-range order in Si-Ge-Sn alloys. 

We speculate that multihyperuniform structures with suppressed large-scale composition fluctuations should be associated with a smaller degree of deviation from Vegard's law as well as smaller lattice distortion that is intrinsic to any HEA or MEA system \cite{latticedistortion} compared to random/SQS structures. We optimize the structures using DFT simulations and determine the lattice constants of the optimized multihyperuniform, random, and SQS structures with $3\times3\times3$ supercells to be 17.903, 17.883, and 17.883~\AA, respectively. According to Vegard's law, the lattice constant of an equi-molar Si-Ge-Sn alloy is predicted to be 17.904~\AA, the weighted average of the lattice constants of pure Si (16.406~\AA), Ge (17.346~\AA), and Sn (19.960~\AA) crystals with $3\times3\times3$ supercells. These results confirm our hypothesis that the multihyperuniform system exhibits an minimal amount of deviation from Vegard's law and reflects the nearly ideal mixing of the three constituent elements in the multihyperuniform SiGeSn MEA. We also compute the lattice distortion using the vector dissimilarity metric \cite{zhuang2021sudoku}, i.e., the root mean squared displacement (RMSD) between the reference lattice (a diamond lattice with a lattice constant as predicted by Vegard's law) and the optimized structure for the multihyperuniform, random, and SQS systems, respectively. We find the RMSDs for the multihyperuniform, random, and SQS systems are 2.14, 3.21, and 3.31 \AA, respectively. Clearly, the multihyperuniform system shows the least extent of lattice distortion, which leads to the lowest ground-state energy of such system. The stability of the multihyperuniform system can be attributed to the suppressed composition fluctuations in such system. On the other hand, the RMSDs and energies for random and SQS systems are nearly the same, as shown in Figure 3(a), which is consistent of the fact that the individual SQSs by design should retain the salient ensemble statistics of random mixtures.


\begin{figure}[ht]
\centering
\includegraphics[width=0.48\textwidth,keepaspectratio]{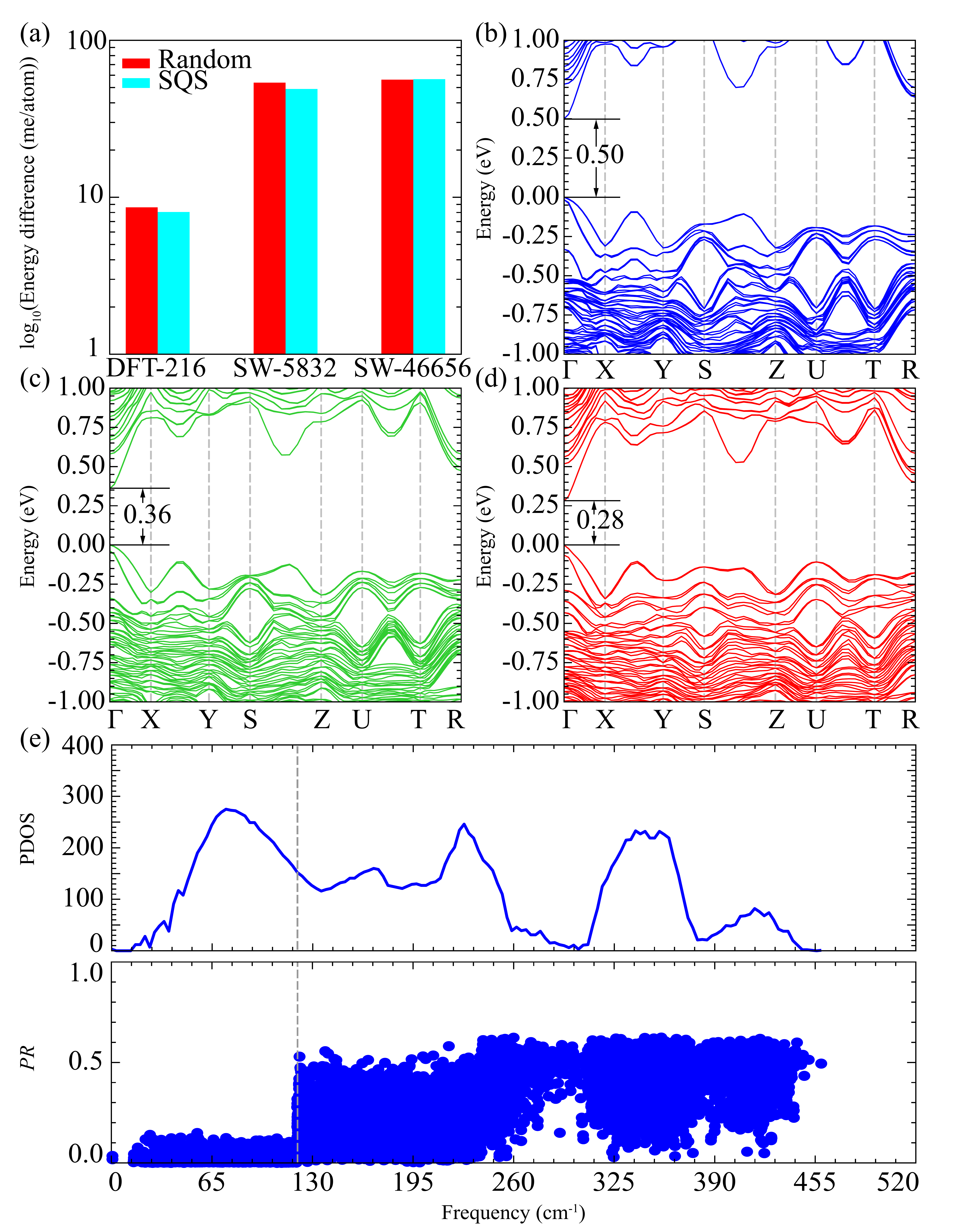}
\caption{(a) Histogram comparing the energies (on a log10 scale) of random, SQS, and multihyperuniform structures using DFT and MD simulations with the Stillinger-Weber potential \cite{PhysRevB.31.5262} and cell sizes. Here we use the multihyperuniform structure as the reference and set its energy to zero; in the plot ``DFT-216'' refers to DFT results with a supercell of 216 atoms, and ``SW-5832'' and ``SW-46656'' refer to MD results with a supercell of 5832 and 46656 atoms, respectively.  Band structures of the multihyperuniform (b), SQS (c), and random (d) \cite{WANG2020443} SiGeSn MEAs. (e) Phonon density of states and partition ratio of SiGeSn MEA with the multihyperuniform structure.}
\end{figure}

{\bf Larger electronic bandgaps.} The effect of the presence of the long-range order on the electronic structure and thermal transport properties of the SiGeSn MEA remains elusive, which we address in our current work. We first calculate the band structure of the SiGeSn MEA with the multihyperuniform and SQS structure and use the previously computed band structure of the random structures for comparison \cite{WANG2020443}. Figures 3(c), (d), and (e) reveals that all the three systems have direct band gaps at the $\Gamma$ point and the corresponding band gaps increase from 0.28, 0.36, to 0.50 eV for the random, SQS, and multihyperuniform systems, respectively. For pure crystalline Si, Ge, and Sn, the band gaps are 1.29, 0.33, and 0.15 eV, respectively. In the random SiGeSn system, the clustering of Ge and Sn atoms significantly reduces the band gap and the presence of MHLRO redistributes all atoms with suppressed local clustering of atoms of the same type, leading to a band gap that is nearly the average band gap (0.59 eV) of the three constituent elements, which is another manifestation that the MHLRO SiGeSn alloy approximately realizes Vegard's law.  

{\bf Enhanced low-temperature thermal conductivity.} The thermal conductivity of the multihyperuniform system at low temperatures are generally higher than that of the random mixture. For example, the thermal conductivities are 1.53 and 0.87 W/m$\cdot$ K at 10 K for the multihyperuniform and random systems, respectively. The improved thermal conductivity makes the multihyperuniform SiGeSn MEA a better candidate for electronic device applications such as computer chips that require a stricter management in heat dissipation. The higher thermal conductivity of the multihyperuniform system once again can be attributed to the smaller lattice distortions of such system compared to the random mixture, since it is well known that in solid solutions such as alloys, large lattice distortions would enhance the scattering of phonons and lead to low thermal conductivity \cite{Fa16}. However, as the temperature increases, the effect of phonon-phonon coupling becomes more prominent, which leads to similar and decreased thermal conductivities for multihyperuniform and random mixtures. For example, we find that the computed thermal conductivities at 400 K for the multihyperuniform and random systems are 0.53 and 0.57 W/m$\cdot$ K, respectively. Figure 3(e) shows the phonon density of states (PDOS) and partition ratio (PR) of the multihyperuniform mixture. Although the overall shape of the PDOS is similar to that of the random system \cite{WANG2020443}, we observe a gap almost opens around 280 cm$^{-1}$ (corresponding to a temperature of about 400 K), leading to diminishing PR modes in the vicinity of this frequency, and a reduction in the thermal conductivity.


In summary, we present strong numerical evidence suggesting a new universal MHLRO in MEA and HEA systems, exemplified by the SiGeSn MEA that is endowed with a minimal lattice distortion and enhanced electronic band gap, both of which matching the predictions from Vegard's law that reflects the ideal mixing behavior of atoms in the MEAs. We also find improved thermal conductivity of the multihyperuniform alloy at low temperatures due to smaller lattice distortions compared to random mixtures. Our analysis of the SiGeSn system leads to the formulation of general organizing principles applicable in other MEAs and HEAs, and a new type of highly efficient computational model for rendering realistic large-scale configurations of MEAs and HEAs beyond the conventional random mixture/SQS models. Our results will open up many novel potential applications in optoelectronics, thermoelectrics, and quantum devices.

\begin{acknowledgments}
X. J., D. W., and H.Z. thank Arizona State University (ASU) for the start-up funds. This research used computational resources of the Agave Research Computer Cluster of ASU and the Texas Advanced Computing Center under Contract TG-DMR170070.
\end{acknowledgments}


\begin{thebibliography}{52}
\expandafter\ifx\csname natexlab\endcsname\relax\def\natexlab#1{#1}\fi
\expandafter\ifx\csname bibnamefont\endcsname\relax
  \def\bibnamefont#1{#1}\fi
\expandafter\ifx\csname bibfnamefont\endcsname\relax
  \def\bibfnamefont#1{#1}\fi
\expandafter\ifx\csname citenamefont\endcsname\relax
  \def\citenamefont#1{#1}\fi
\expandafter\ifx\csname url\endcsname\relax
  \def\url#1{\texttt{#1}}\fi
\expandafter\ifx\csname urlprefix\endcsname\relax\def\urlprefix{URL }\fi
\providecommand{\bibinfo}[2]{#2}
\providecommand{\eprint}[2][]{\url{#2}}

\bibitem[{\citenamefont{Zhang et~al.}(2014)\citenamefont{Zhang, Zuo, Tang, Gao,
  Dahmen, Liaw, and Lu}}]{ZHANG20141}
\bibinfo{author}{\bibfnamefont{Y.}~\bibnamefont{Zhang}},
  \bibinfo{author}{\bibfnamefont{T.~T.} \bibnamefont{Zuo}},
  \bibinfo{author}{\bibfnamefont{Z.}~\bibnamefont{Tang}},
  \bibinfo{author}{\bibfnamefont{M.~C.} \bibnamefont{Gao}},
  \bibinfo{author}{\bibfnamefont{K.~A.} \bibnamefont{Dahmen}},
  \bibinfo{author}{\bibfnamefont{P.~K.} \bibnamefont{Liaw}}, \bibnamefont{and}
  \bibinfo{author}{\bibfnamefont{Z.~P.} \bibnamefont{Lu}},
  \bibinfo{journal}{Prog. Mater. Sci.} \textbf{\bibinfo{volume}{61}},
  \bibinfo{pages}{1} (\bibinfo{year}{2014}).

\bibitem[{\citenamefont{Miracle and Senkov}(2017)}]{MIRACLE2017448}
\bibinfo{author}{\bibfnamefont{D.}~\bibnamefont{Miracle}} \bibnamefont{and}
  \bibinfo{author}{\bibfnamefont{O.}~\bibnamefont{Senkov}},
  \bibinfo{journal}{Acta Mater.} \textbf{\bibinfo{volume}{122}},
  \bibinfo{pages}{448} (\bibinfo{year}{2017}).

\bibitem[{\citenamefont{Mu et~al.}(2018)\citenamefont{Mu, Pei, Liu, and
  Stocks}}]{mu_pei_liu_stocks_2018}
\bibinfo{author}{\bibfnamefont{S.}~\bibnamefont{Mu}},
  \bibinfo{author}{\bibfnamefont{Z.}~\bibnamefont{Pei}},
  \bibinfo{author}{\bibfnamefont{X.}~\bibnamefont{Liu}}, \bibnamefont{and}
  \bibinfo{author}{\bibfnamefont{G.~M.} \bibnamefont{Stocks}},
  \bibinfo{journal}{J. Mater. Sci.} \textbf{\bibinfo{volume}{33}},
  \bibinfo{pages}{2857â€“2880} (\bibinfo{year}{2018}).

\bibitem[{\citenamefont{Zunger et~al.}(1990)\citenamefont{Zunger, Wei,
  Ferreira, and Bernard}}]{PhysRevLett.65.353}
\bibinfo{author}{\bibfnamefont{A.}~\bibnamefont{Zunger}},
  \bibinfo{author}{\bibfnamefont{S.-H.} \bibnamefont{Wei}},
  \bibinfo{author}{\bibfnamefont{L.~G.} \bibnamefont{Ferreira}},
  \bibnamefont{and} \bibinfo{author}{\bibfnamefont{J.~E.}
  \bibnamefont{Bernard}}, \bibinfo{journal}{Phys. Rev. Lett.}
  \textbf{\bibinfo{volume}{65}}, \bibinfo{pages}{353} (\bibinfo{year}{1990}).

\bibitem[{\citenamefont{Ding et~al.}(2018)\citenamefont{Ding, Yu, Asta, and
  Ritchie}}]{ding2018tunable}
\bibinfo{author}{\bibfnamefont{J.}~\bibnamefont{Ding}},
  \bibinfo{author}{\bibfnamefont{Q.}~\bibnamefont{Yu}},
  \bibinfo{author}{\bibfnamefont{M.}~\bibnamefont{Asta}}, \bibnamefont{and}
  \bibinfo{author}{\bibfnamefont{R.~O.} \bibnamefont{Ritchie}},
  \bibinfo{journal}{Proc. Natl. Acad. Sci. U.S.A.}
  \textbf{\bibinfo{volume}{115}}, \bibinfo{pages}{8919} (\bibinfo{year}{2018}).

\bibitem[{\citenamefont{Zhang et~al.}(2020)\citenamefont{Zhang, Zhao, Ding,
  Chong, Jia, Ophus, Asta, Ritchie, and Minor}}]{zhang2020short}
\bibinfo{author}{\bibfnamefont{R.}~\bibnamefont{Zhang}},
  \bibinfo{author}{\bibfnamefont{S.}~\bibnamefont{Zhao}},
  \bibinfo{author}{\bibfnamefont{J.}~\bibnamefont{Ding}},
  \bibinfo{author}{\bibfnamefont{Y.}~\bibnamefont{Chong}},
  \bibinfo{author}{\bibfnamefont{T.}~\bibnamefont{Jia}},
  \bibinfo{author}{\bibfnamefont{C.}~\bibnamefont{Ophus}},
  \bibinfo{author}{\bibfnamefont{M.}~\bibnamefont{Asta}},
  \bibinfo{author}{\bibfnamefont{R.~O.} \bibnamefont{Ritchie}},
  \bibnamefont{and} \bibinfo{author}{\bibfnamefont{A.~M.} \bibnamefont{Minor}},
  \bibinfo{journal}{Nature} \textbf{\bibinfo{volume}{581}},
  \bibinfo{pages}{283} (\bibinfo{year}{2020}).

\bibitem[{\citenamefont{Walsh et~al.}(2021)\citenamefont{Walsh, Asta, and
  Ritchie}}]{walsh2021magnetically}
\bibinfo{author}{\bibfnamefont{F.}~\bibnamefont{Walsh}},
  \bibinfo{author}{\bibfnamefont{M.}~\bibnamefont{Asta}}, \bibnamefont{and}
  \bibinfo{author}{\bibfnamefont{R.~O.} \bibnamefont{Ritchie}},
  \bibinfo{journal}{Proc. Natl. Acad. Sci. U.S.A.}
  \textbf{\bibinfo{volume}{118}} (\bibinfo{year}{2021}).

\bibitem[{\citenamefont{Torquato and Stillinger}(2003)}]{To03}
\bibinfo{author}{\bibfnamefont{S.}~\bibnamefont{Torquato}} \bibnamefont{and}
  \bibinfo{author}{\bibfnamefont{F.~H.} \bibnamefont{Stillinger}},
  \bibinfo{journal}{Phys. Rev. E} \textbf{\bibinfo{volume}{68}},
  \bibinfo{pages}{041113} (\bibinfo{year}{2003}).

\bibitem[{\citenamefont{Torquato}(2018)}]{To18a}
\bibinfo{author}{\bibfnamefont{S.}~\bibnamefont{Torquato}},
  \bibinfo{journal}{Phys. Rep.} \textbf{\bibinfo{volume}{745}},
  \bibinfo{pages}{1} (\bibinfo{year}{2018}).

\bibitem[{\citenamefont{Donev et~al.}(2005)\citenamefont{Donev, Stillinger, and
  Torquato}}]{Do05}
\bibinfo{author}{\bibfnamefont{A.}~\bibnamefont{Donev}},
  \bibinfo{author}{\bibfnamefont{F.~H.} \bibnamefont{Stillinger}},
  \bibnamefont{and} \bibinfo{author}{\bibfnamefont{S.}~\bibnamefont{Torquato}},
  \bibinfo{journal}{Phys. Rev. Lett.} \textbf{\bibinfo{volume}{95}},
  \bibinfo{pages}{090604} (\bibinfo{year}{2005}).

\bibitem[{\citenamefont{Zachary et~al.}(2011)\citenamefont{Zachary, Jiao, and
  Torquato}}]{Za11a}
\bibinfo{author}{\bibfnamefont{C.~E.} \bibnamefont{Zachary}},
  \bibinfo{author}{\bibfnamefont{Y.}~\bibnamefont{Jiao}}, \bibnamefont{and}
  \bibinfo{author}{\bibfnamefont{S.}~\bibnamefont{Torquato}},
  \bibinfo{journal}{Phys. Rev. Lett.} \textbf{\bibinfo{volume}{106}},
  \bibinfo{pages}{178001} (\bibinfo{year}{2011}).

\bibitem[{\citenamefont{Jiao and Torquato}(2011)}]{Ji11}
\bibinfo{author}{\bibfnamefont{Y.}~\bibnamefont{Jiao}} \bibnamefont{and}
  \bibinfo{author}{\bibfnamefont{S.}~\bibnamefont{Torquato}},
  \bibinfo{journal}{Phys. Rev. E} \textbf{\bibinfo{volume}{84}},
  \bibinfo{pages}{041309} (\bibinfo{year}{2011}).

\bibitem[{\citenamefont{Jiao et~al.}(2014)\citenamefont{Jiao, Lau, Hatzikirou,
  Meyer-Hermann, Corbo, and Torquato}}]{Ji14}
\bibinfo{author}{\bibfnamefont{Y.}~\bibnamefont{Jiao}},
  \bibinfo{author}{\bibfnamefont{T.}~\bibnamefont{Lau}},
  \bibinfo{author}{\bibfnamefont{H.}~\bibnamefont{Hatzikirou}},
  \bibinfo{author}{\bibfnamefont{M.}~\bibnamefont{Meyer-Hermann}},
  \bibinfo{author}{\bibfnamefont{J.~C.} \bibnamefont{Corbo}}, \bibnamefont{and}
  \bibinfo{author}{\bibfnamefont{S.}~\bibnamefont{Torquato}},
  \bibinfo{journal}{Phys. Rev. E} \textbf{\bibinfo{volume}{89}},
  \bibinfo{pages}{022721} (\bibinfo{year}{2014}).

\bibitem[{\citenamefont{Zachary and Torquato}(2009)}]{Za09}
\bibinfo{author}{\bibfnamefont{C.~E.} \bibnamefont{Zachary}} \bibnamefont{and}
  \bibinfo{author}{\bibfnamefont{S.}~\bibnamefont{Torquato}},
  \bibinfo{journal}{J. Stat. Mech. Theor. Exp.}
  \textbf{\bibinfo{volume}{2009}}, \bibinfo{pages}{P12015}
  (\bibinfo{year}{2009}).

\bibitem[{\citenamefont{Chen et~al.}(2014)\citenamefont{Chen, Jiao, and
  Torquato}}]{Ch14}
\bibinfo{author}{\bibfnamefont{D.}~\bibnamefont{Chen}},
  \bibinfo{author}{\bibfnamefont{Y.}~\bibnamefont{Jiao}}, \bibnamefont{and}
  \bibinfo{author}{\bibfnamefont{S.}~\bibnamefont{Torquato}},
  \bibinfo{journal}{J. Phys. Chem. B} \textbf{\bibinfo{volume}{118}},
  \bibinfo{pages}{7981} (\bibinfo{year}{2014}).

\bibitem[{\citenamefont{Zachary and Torquato}(2011)}]{Za11b}
\bibinfo{author}{\bibfnamefont{C.~E.} \bibnamefont{Zachary}} \bibnamefont{and}
  \bibinfo{author}{\bibfnamefont{S.}~\bibnamefont{Torquato}},
  \bibinfo{journal}{Phys. Rev. E} \textbf{\bibinfo{volume}{83}},
  \bibinfo{pages}{051133} (\bibinfo{year}{2011}).

\bibitem[{\citenamefont{Torquato et~al.}(2015)\citenamefont{Torquato, Zhang,
  and Stillinger}}]{To15}
\bibinfo{author}{\bibfnamefont{S.}~\bibnamefont{Torquato}},
  \bibinfo{author}{\bibfnamefont{G.}~\bibnamefont{Zhang}}, \bibnamefont{and}
  \bibinfo{author}{\bibfnamefont{F.~H.} \bibnamefont{Stillinger}},
  \bibinfo{journal}{Phys. Rev. X} \textbf{\bibinfo{volume}{5}},
  \bibinfo{pages}{021020} (\bibinfo{year}{2015}).

\bibitem[{\citenamefont{Batten et~al.}(2009)\citenamefont{Batten, Stillinger,
  and Torquato}}]{Ba09}
\bibinfo{author}{\bibfnamefont{R.~D.} \bibnamefont{Batten}},
  \bibinfo{author}{\bibfnamefont{F.~H.} \bibnamefont{Stillinger}},
  \bibnamefont{and} \bibinfo{author}{\bibfnamefont{S.}~\bibnamefont{Torquato}},
  \bibinfo{journal}{Phys. Rev. Lett.} \textbf{\bibinfo{volume}{103}},
  \bibinfo{pages}{050602} (\bibinfo{year}{2009}).

\bibitem[{\citenamefont{Kurita and Weeks}(2011)}]{Ku11}
\bibinfo{author}{\bibfnamefont{R.}~\bibnamefont{Kurita}} \bibnamefont{and}
  \bibinfo{author}{\bibfnamefont{E.~R.} \bibnamefont{Weeks}},
  \bibinfo{journal}{Phys. Rev. E} \textbf{\bibinfo{volume}{84}},
  \bibinfo{pages}{030401(R)} (\bibinfo{year}{2011}).

\bibitem[{\citenamefont{Hunter and Weeks}(2012)}]{Hu122}
\bibinfo{author}{\bibfnamefont{G.~L.} \bibnamefont{Hunter}} \bibnamefont{and}
  \bibinfo{author}{\bibfnamefont{E.~R.} \bibnamefont{Weeks}},
  \bibinfo{journal}{Rep. Prog. Phys.} \textbf{\bibinfo{volume}{75}},
  \bibinfo{pages}{066501} (\bibinfo{year}{2012}).

\bibitem[{\citenamefont{Dreyfus et~al.}(2015)\citenamefont{Dreyfus, Xu, Still,
  Hough, Yodh, and Torquato}}]{Dr15}
\bibinfo{author}{\bibfnamefont{R.}~\bibnamefont{Dreyfus}},
  \bibinfo{author}{\bibfnamefont{Y.}~\bibnamefont{Xu}},
  \bibinfo{author}{\bibfnamefont{T.}~\bibnamefont{Still}},
  \bibinfo{author}{\bibfnamefont{L.~A.} \bibnamefont{Hough}},
  \bibinfo{author}{\bibfnamefont{A.~G.} \bibnamefont{Yodh}}, \bibnamefont{and}
  \bibinfo{author}{\bibfnamefont{S.}~\bibnamefont{Torquato}},
  \bibinfo{journal}{Phys. Rev. E} \textbf{\bibinfo{volume}{91}},
  \bibinfo{pages}{012302} (\bibinfo{year}{2015}).

\bibitem[{\citenamefont{Hexner and Levine}(2015)}]{He15}
\bibinfo{author}{\bibfnamefont{D.}~\bibnamefont{Hexner}} \bibnamefont{and}
  \bibinfo{author}{\bibfnamefont{D.}~\bibnamefont{Levine}},
  \bibinfo{journal}{Phys. Rev. Lett.} \textbf{\bibinfo{volume}{114}},
  \bibinfo{pages}{110602} (\bibinfo{year}{2015}).

\bibitem[{\citenamefont{Jack et~al.}(2015)\citenamefont{Jack, Thompson, and
  Sollich}}]{Ja15}
\bibinfo{author}{\bibfnamefont{R.~L.} \bibnamefont{Jack}},
  \bibinfo{author}{\bibfnamefont{I.~R.} \bibnamefont{Thompson}},
  \bibnamefont{and} \bibinfo{author}{\bibfnamefont{P.}~\bibnamefont{Sollich}},
  \bibinfo{journal}{Phys. Rev. Lett.} \textbf{\bibinfo{volume}{114}},
  \bibinfo{pages}{060601} (\bibinfo{year}{2015}).

\bibitem[{\citenamefont{Weijs et~al.}(2015)\citenamefont{Weijs, Jeanneret,
  Dreyfus, and Bartolo}}]{We15}
\bibinfo{author}{\bibfnamefont{J.~H.} \bibnamefont{Weijs}},
  \bibinfo{author}{\bibfnamefont{R.}~\bibnamefont{Jeanneret}},
  \bibinfo{author}{\bibfnamefont{R.}~\bibnamefont{Dreyfus}}, \bibnamefont{and}
  \bibinfo{author}{\bibfnamefont{D.}~\bibnamefont{Bartolo}},
  \bibinfo{journal}{Phys. Rev. Lett.} \textbf{\bibinfo{volume}{115}},
  \bibinfo{pages}{108301} (\bibinfo{year}{2015}).

\bibitem[{\citenamefont{Florescu et~al.}(2009)\citenamefont{Florescu, Torquato,
  and Steinhardt}}]{Fl09}
\bibinfo{author}{\bibfnamefont{M.}~\bibnamefont{Florescu}},
  \bibinfo{author}{\bibfnamefont{S.}~\bibnamefont{Torquato}}, \bibnamefont{and}
  \bibinfo{author}{\bibfnamefont{P.~J.} \bibnamefont{Steinhardt}},
  \bibinfo{journal}{Proc. Natl. Acad. Sci. U.S.A.}
  \textbf{\bibinfo{volume}{106}}, \bibinfo{pages}{20658}
  (\bibinfo{year}{2009}).

\bibitem[{\citenamefont{Man et~al.}(2013)\citenamefont{Man, Florescu,
  Williamson, He, Hashemizad, Leung, Liner, Torquato, Chaikin, and
  Steinhardt}}]{Ma13}
\bibinfo{author}{\bibfnamefont{W.}~\bibnamefont{Man}},
  \bibinfo{author}{\bibfnamefont{M.}~\bibnamefont{Florescu}},
  \bibinfo{author}{\bibfnamefont{E.~P.} \bibnamefont{Williamson}},
  \bibinfo{author}{\bibfnamefont{Y.}~\bibnamefont{He}},
  \bibinfo{author}{\bibfnamefont{S.~R.} \bibnamefont{Hashemizad}},
  \bibinfo{author}{\bibfnamefont{B.~Y.~C.} \bibnamefont{Leung}},
  \bibinfo{author}{\bibfnamefont{D.~R.} \bibnamefont{Liner}},
  \bibinfo{author}{\bibfnamefont{S.}~\bibnamefont{Torquato}},
  \bibinfo{author}{\bibfnamefont{P.~M.} \bibnamefont{Chaikin}},
  \bibnamefont{and} \bibinfo{author}{\bibfnamefont{P.~J.}
  \bibnamefont{Steinhardt}}, \bibinfo{journal}{Proc. Natl. Acad. Sci. U.S.A.}
  \textbf{\bibinfo{volume}{110}}, \bibinfo{pages}{15886}
  (\bibinfo{year}{2013}).

\bibitem[{\citenamefont{Mayer et~al.}(2015)\citenamefont{Mayer,
  Balasubramanian, Mora, and Walczak}}]{Ma15}
\bibinfo{author}{\bibfnamefont{A.}~\bibnamefont{Mayer}},
  \bibinfo{author}{\bibfnamefont{V.}~\bibnamefont{Balasubramanian}},
  \bibinfo{author}{\bibfnamefont{T.}~\bibnamefont{Mora}}, \bibnamefont{and}
  \bibinfo{author}{\bibfnamefont{A.~M.} \bibnamefont{Walczak}},
  \bibinfo{journal}{Proc. Natl. Acad. Sci. USA} \textbf{\bibinfo{volume}{112}},
  \bibinfo{pages}{5950} (\bibinfo{year}{2015}).

\bibitem[{\citenamefont{Hejna et~al.}(2013)\citenamefont{Hejna, Steinhardt, and
  Torquato}}]{He13}
\bibinfo{author}{\bibfnamefont{M.}~\bibnamefont{Hejna}},
  \bibinfo{author}{\bibfnamefont{P.~J.} \bibnamefont{Steinhardt}},
  \bibnamefont{and} \bibinfo{author}{\bibfnamefont{S.}~\bibnamefont{Torquato}},
  \bibinfo{journal}{Phys. Rev. B} \textbf{\bibinfo{volume}{87}},
  \bibinfo{pages}{245204} (\bibinfo{year}{2013}).

\bibitem[{\citenamefont{Klatt et~al.}(2019)\citenamefont{Klatt, Lovri{\'c},
  Chen, Kapfer, Schaller, Sch{\"o}nh{\"o}fer, Gardiner, Smith,
  Schr{\"o}der-Turk, and Torquato}}]{Kl19}
\bibinfo{author}{\bibfnamefont{M.~A.} \bibnamefont{Klatt}},
  \bibinfo{author}{\bibfnamefont{J.}~\bibnamefont{Lovri{\'c}}},
  \bibinfo{author}{\bibfnamefont{D.}~\bibnamefont{Chen}},
  \bibinfo{author}{\bibfnamefont{S.~C.} \bibnamefont{Kapfer}},
  \bibinfo{author}{\bibfnamefont{F.~M.} \bibnamefont{Schaller}},
  \bibinfo{author}{\bibfnamefont{P.~W.~A.} \bibnamefont{Sch{\"o}nh{\"o}fer}},
  \bibinfo{author}{\bibfnamefont{B.~S.} \bibnamefont{Gardiner}},
  \bibinfo{author}{\bibfnamefont{A.}~\bibnamefont{Smith}},
  \bibinfo{author}{\bibfnamefont{G.~E.} \bibnamefont{Schr{\"o}der-Turk}},
  \bibnamefont{and} \bibinfo{author}{\bibfnamefont{S.}~\bibnamefont{Torquato}},
  \bibinfo{journal}{Nat. Commun.} \textbf{\bibinfo{volume}{10}},
  \bibinfo{pages}{1} (\bibinfo{year}{2019}).

\bibitem[{\citenamefont{Lei et~al.}(2019)\citenamefont{Lei, Ciamarra, and
  Ni}}]{Le19a}
\bibinfo{author}{\bibfnamefont{Q.-L.} \bibnamefont{Lei}},
  \bibinfo{author}{\bibfnamefont{M.~P.} \bibnamefont{Ciamarra}},
  \bibnamefont{and} \bibinfo{author}{\bibfnamefont{R.}~\bibnamefont{Ni}},
  \bibinfo{journal}{Sci. Adv.} \textbf{\bibinfo{volume}{5}},
  \bibinfo{pages}{eaau7423} (\bibinfo{year}{2019}).

\bibitem[{\citenamefont{Lei and Ni}(2019)}]{Le19b}
\bibinfo{author}{\bibfnamefont{Q.-L.} \bibnamefont{Lei}} \bibnamefont{and}
  \bibinfo{author}{\bibfnamefont{R.}~\bibnamefont{Ni}}, \bibinfo{journal}{Proc.
  Natl. Acad. Sci. U.S.A.} \textbf{\bibinfo{volume}{116}},
  \bibinfo{pages}{22983} (\bibinfo{year}{2019}).

\bibitem[{\citenamefont{Chen and Torquato}(2018)}]{Ch18a}
\bibinfo{author}{\bibfnamefont{D.}~\bibnamefont{Chen}} \bibnamefont{and}
  \bibinfo{author}{\bibfnamefont{S.}~\bibnamefont{Torquato}},
  \bibinfo{journal}{Acta Mater.} \textbf{\bibinfo{volume}{142}},
  \bibinfo{pages}{152} (\bibinfo{year}{2018}).

\bibitem[{\citenamefont{Chremos and Douglas}(2018)}]{Ch18b}
\bibinfo{author}{\bibfnamefont{A.}~\bibnamefont{Chremos}} \bibnamefont{and}
  \bibinfo{author}{\bibfnamefont{J.~F.} \bibnamefont{Douglas}},
  \bibinfo{journal}{Phys. Rev. Lett.} \textbf{\bibinfo{volume}{121}},
  \bibinfo{pages}{258002} (\bibinfo{year}{2018}).

\bibitem[{\citenamefont{Rumi et~al.}(2019)\citenamefont{Rumi, S{\'a}nchez,
  El{\'\i}as, Maldonado, Puig, Bolecek, Nieva, Konczykowski, Fasano, and
  Kolton}}]{Ru19}
\bibinfo{author}{\bibfnamefont{G.}~\bibnamefont{Rumi}},
  \bibinfo{author}{\bibfnamefont{J.~A.} \bibnamefont{S{\'a}nchez}},
  \bibinfo{author}{\bibfnamefont{F.}~\bibnamefont{El{\'\i}as}},
  \bibinfo{author}{\bibfnamefont{R.~C.} \bibnamefont{Maldonado}},
  \bibinfo{author}{\bibfnamefont{J.}~\bibnamefont{Puig}},
  \bibinfo{author}{\bibfnamefont{N.~R.~C.} \bibnamefont{Bolecek}},
  \bibinfo{author}{\bibfnamefont{G.}~\bibnamefont{Nieva}},
  \bibinfo{author}{\bibfnamefont{M.}~\bibnamefont{Konczykowski}},
  \bibinfo{author}{\bibfnamefont{Y.}~\bibnamefont{Fasano}}, \bibnamefont{and}
  \bibinfo{author}{\bibfnamefont{A.~B.} \bibnamefont{Kolton}},
  \bibinfo{journal}{Phys. Rev. Res.} \textbf{\bibinfo{volume}{1}},
  \bibinfo{pages}{033057} (\bibinfo{year}{2019}).

\bibitem[{\citenamefont{Huang et~al.}(2021)\citenamefont{Huang, Hu, Yang, Liu,
  and Zhang}}]{Hu21}
\bibinfo{author}{\bibfnamefont{M.}~\bibnamefont{Huang}},
  \bibinfo{author}{\bibfnamefont{W.}~\bibnamefont{Hu}},
  \bibinfo{author}{\bibfnamefont{S.}~\bibnamefont{Yang}},
  \bibinfo{author}{\bibfnamefont{Q.-X.} \bibnamefont{Liu}}, \bibnamefont{and}
  \bibinfo{author}{\bibfnamefont{H.~P.} \bibnamefont{Zhang}},
  \bibinfo{journal}{Proc. Natl. Acad. Sci. U.S.A.}
  \textbf{\bibinfo{volume}{118}}, \bibinfo{pages}{e2100493118}
  (\bibinfo{year}{2021}).

\bibitem[{\citenamefont{Torquato}(2021)}]{To21}
\bibinfo{author}{\bibfnamefont{S.}~\bibnamefont{Torquato}},
  \bibinfo{journal}{Proc. Natl. Acad. Sci. U.S.A.}
  \textbf{\bibinfo{volume}{118}}, \bibinfo{pages}{e2107276118}
  (\bibinfo{year}{2021}).

\bibitem[{\citenamefont{Jiao}(2022)}]{jiao2021hyperuniformity}
\bibinfo{author}{\bibfnamefont{Y.}~\bibnamefont{Jiao}}, \bibinfo{journal}{Phys.
  A: Stat. Mech. Appl.} \textbf{\bibinfo{volume}{585}}, \bibinfo{pages}{126435}
  (\bibinfo{year}{2022}).

\bibitem[{\citenamefont{Vegard}(1921)}]{vegard1921konstitution}
\bibinfo{author}{\bibfnamefont{L.}~\bibnamefont{Vegard}}, \bibinfo{journal}{Z.
  Phys.} \textbf{\bibinfo{volume}{5}}, \bibinfo{pages}{17}
  (\bibinfo{year}{1921}).

\bibitem[{\citenamefont{Cowley}(1950)}]{Co50}
\bibinfo{author}{\bibfnamefont{J.~M.} \bibnamefont{Cowley}},
  \bibinfo{journal}{Phys. Rev.} \textbf{\bibinfo{volume}{77}},
  \bibinfo{pages}{669} (\bibinfo{year}{1950}).

\bibitem[{\citenamefont{Olesinski and
  Abbaschian}(1984{\natexlab{a}})}]{olesinski1984si}
\bibinfo{author}{\bibfnamefont{R.}~\bibnamefont{Olesinski}} \bibnamefont{and}
  \bibinfo{author}{\bibfnamefont{G.}~\bibnamefont{Abbaschian}},
  \bibinfo{journal}{Bull. Alloy Phase Diagr.} \textbf{\bibinfo{volume}{5}},
  \bibinfo{pages}{273} (\bibinfo{year}{1984}{\natexlab{a}}).

\bibitem[{\citenamefont{Olesinski and
  Abbaschian}(1984{\natexlab{b}})}]{olesinski1984ge1}
\bibinfo{author}{\bibfnamefont{R.}~\bibnamefont{Olesinski}} \bibnamefont{and}
  \bibinfo{author}{\bibfnamefont{G.}~\bibnamefont{Abbaschian}},
  \bibinfo{journal}{Bull. Alloy Phase Diagr.} \textbf{\bibinfo{volume}{5}},
  \bibinfo{pages}{265} (\bibinfo{year}{1984}{\natexlab{b}}).

\bibitem[{\citenamefont{Olesinski and
  Abbaschian}(1984{\natexlab{c}})}]{olesinski1984ge2}
\bibinfo{author}{\bibfnamefont{R.}~\bibnamefont{Olesinski}} \bibnamefont{and}
  \bibinfo{author}{\bibfnamefont{G.}~\bibnamefont{Abbaschian}},
  \bibinfo{journal}{Bull. Alloy Phase Diagr.} \textbf{\bibinfo{volume}{5}},
  \bibinfo{pages}{180} (\bibinfo{year}{1984}{\natexlab{c}}).

\bibitem[{\citenamefont{Mukherjee et~al.}(2017)\citenamefont{Mukherjee, Kodali,
  Isheim, Wirths, Hartmann, Buca, Seidman, and
  Moutanabbir}}]{PhysRevB.95.161402}
\bibinfo{author}{\bibfnamefont{S.}~\bibnamefont{Mukherjee}},
  \bibinfo{author}{\bibfnamefont{N.}~\bibnamefont{Kodali}},
  \bibinfo{author}{\bibfnamefont{D.}~\bibnamefont{Isheim}},
  \bibinfo{author}{\bibfnamefont{S.}~\bibnamefont{Wirths}},
  \bibinfo{author}{\bibfnamefont{J.~M.} \bibnamefont{Hartmann}},
  \bibinfo{author}{\bibfnamefont{D.}~\bibnamefont{Buca}},
  \bibinfo{author}{\bibfnamefont{D.~N.} \bibnamefont{Seidman}},
  \bibnamefont{and}
  \bibinfo{author}{\bibfnamefont{O.}~\bibnamefont{Moutanabbir}},
  \bibinfo{journal}{Phys. Rev. B} \textbf{\bibinfo{volume}{95}},
  \bibinfo{pages}{161402} (\bibinfo{year}{2017}).

\bibitem[{\citenamefont{Cao et~al.}(2020)\citenamefont{Cao, Chen, Jin, Liu, and
  Li}}]{cao2020short}
\bibinfo{author}{\bibfnamefont{B.}~\bibnamefont{Cao}},
  \bibinfo{author}{\bibfnamefont{S.}~\bibnamefont{Chen}},
  \bibinfo{author}{\bibfnamefont{X.}~\bibnamefont{Jin}},
  \bibinfo{author}{\bibfnamefont{J.}~\bibnamefont{Liu}}, \bibnamefont{and}
  \bibinfo{author}{\bibfnamefont{T.}~\bibnamefont{Li}}, \bibinfo{journal}{ACS
  Appl. Mater. Interfaces} \textbf{\bibinfo{volume}{12}},
  \bibinfo{pages}{57245} (\bibinfo{year}{2020}).

\bibitem[{\citenamefont{Wang et~al.}(2019)\citenamefont{Wang, Liu, Huang, and
  Zhuang}}]{doi:10.1063/1.5135324}
\bibinfo{author}{\bibfnamefont{D.}~\bibnamefont{Wang}},
  \bibinfo{author}{\bibfnamefont{L.}~\bibnamefont{Liu}},
  \bibinfo{author}{\bibfnamefont{W.}~\bibnamefont{Huang}}, \bibnamefont{and}
  \bibinfo{author}{\bibfnamefont{H.~L.} \bibnamefont{Zhuang}},
  \bibinfo{journal}{J. Appl. Phys.} \textbf{\bibinfo{volume}{126}},
  \bibinfo{pages}{225703} (\bibinfo{year}{2019}).

\bibitem[{\citenamefont{Stillinger and Weber}(1985)}]{PhysRevB.31.5262}
\bibinfo{author}{\bibfnamefont{F.~H.} \bibnamefont{Stillinger}}
  \bibnamefont{and} \bibinfo{author}{\bibfnamefont{T.~A.} \bibnamefont{Weber}},
  \bibinfo{journal}{Phys. Rev. B} \textbf{\bibinfo{volume}{31}},
  \bibinfo{pages}{5262} (\bibinfo{year}{1985}).

\bibitem[{\citenamefont{Lee and Hwang}(2012)}]{lee2012force}
\bibinfo{author}{\bibfnamefont{Y.}~\bibnamefont{Lee}} \bibnamefont{and}
  \bibinfo{author}{\bibfnamefont{G.~S.} \bibnamefont{Hwang}},
  \bibinfo{journal}{Phys. Rev. B} \textbf{\bibinfo{volume}{85}},
  \bibinfo{pages}{125204} (\bibinfo{year}{2012}).

\bibitem[{\citenamefont{Lee and Hwang}(2017)}]{lee2017molecular}
\bibinfo{author}{\bibfnamefont{Y.}~\bibnamefont{Lee}} \bibnamefont{and}
  \bibinfo{author}{\bibfnamefont{G.~S.} \bibnamefont{Hwang}},
  \bibinfo{journal}{J. Phys. D: Appl. Phys.}
  \textbf{\bibinfo{volume}{50}}, \bibinfo{pages}{494001}
  (\bibinfo{year}{2017}).

\bibitem[{\citenamefont{Wang et~al.}(2020)\citenamefont{Wang, Liu, Chen, and
  Zhuang}}]{WANG2020443}
\bibinfo{author}{\bibfnamefont{D.}~\bibnamefont{Wang}},
  \bibinfo{author}{\bibfnamefont{L.}~\bibnamefont{Liu}},
  \bibinfo{author}{\bibfnamefont{M.}~\bibnamefont{Chen}}, \bibnamefont{and}
  \bibinfo{author}{\bibfnamefont{H.}~\bibnamefont{Zhuang}},
  \bibinfo{journal}{Acta Mater.} \textbf{\bibinfo{volume}{199}},
  \bibinfo{pages}{443} (\bibinfo{year}{2020}).

\bibitem[{\citenamefont{He and Yang}(2018)}]{latticedistortion}
\bibinfo{author}{\bibfnamefont{Q.}~\bibnamefont{He}} \bibnamefont{and}
  \bibinfo{author}{\bibfnamefont{Y.}~\bibnamefont{Yang}},
  \bibinfo{journal}{Front. Mater. Sci.} \textbf{\bibinfo{volume}{5}},
  \bibinfo{pages}{42} (\bibinfo{year}{2018}).

\bibitem[{\citenamefont{Zhuang}(2021)}]{zhuang2021sudoku}
\bibinfo{author}{\bibfnamefont{H.}~\bibnamefont{Zhuang}},
  \bibinfo{journal}{arXiv preprint arXiv:2110.03797}  (\bibinfo{year}{2021}).

\bibitem[{\citenamefont{Fan et~al.}(2016)\citenamefont{Fan, Wang, Wu, Liu, and
  Lu}}]{Fa16}
\bibinfo{author}{\bibfnamefont{Z.}~\bibnamefont{Fan}},
  \bibinfo{author}{\bibfnamefont{H.}~\bibnamefont{Wang}},
  \bibinfo{author}{\bibfnamefont{Y.}~\bibnamefont{Wu}},
  \bibinfo{author}{\bibfnamefont{X.~J.} \bibnamefont{Liu}}, \bibnamefont{and}
  \bibinfo{author}{\bibfnamefont{Z.~P.} \bibnamefont{Lu}},
  \bibinfo{journal}{Rsc Adv.} \textbf{\bibinfo{volume}{6}},
  \bibinfo{pages}{52164} (\bibinfo{year}{2016}).

\end{thebibliography}

\end{document}